\documentclass[twoside,twocolumn,9pt]{article}
\usepackage{extsizes}
\usepackage[super,sort&compress,comma]{natbib} 
\usepackage[version=3]{mhchem}
\usepackage[left=1.5cm, right=1.5cm, top=1.785cm, bottom=2.0cm]{geometry}
\usepackage{balance}
\usepackage{mathptmx}
\usepackage{sectsty}
\usepackage{graphicx} 
\usepackage{lastpage}
\usepackage[format=plain,justification=justified,singlelinecheck=false,font={stretch=1.125,small,sf},labelfont=bf,labelsep=space]{caption}
\usepackage{float}
\usepackage{fancyhdr}
\usepackage{fnpos}
\usepackage[english]{babel}
\addto{\captionsenglish}{%
  
}
\usepackage{array}
\usepackage{droidsans}
\usepackage{charter}
\usepackage[T1]{fontenc}
\usepackage[usenames,dvipsnames]{xcolor}
\usepackage{setspace}
\usepackage[compact]{titlesec}
\usepackage{hyperref}

\usepackage{stmaryrd}
\usepackage{epstopdf}
\definecolor{cream}{RGB}{222,217,201}

\begin{document}

\pagestyle{fancy}
\thispagestyle{plain}
\fancypagestyle{plain}{
\renewcommand{\headrulewidth}{0pt}
}

\makeFNbottom
\makeatletter
\renewcommand\LARGE{\@setfontsize\LARGE{15pt}{17}}
\renewcommand\Large{\@setfontsize\Large{12pt}{14}}
\renewcommand\large{\@setfontsize\large{10pt}{12}}
\renewcommand\footnotesize{\@setfontsize\footnotesize{7pt}{10}}
\renewcommand\scriptsize{\@setfontsize\scriptsize{7pt}{7}}
\makeatother

\renewcommand{\thefootnote}{\fnsymbol{footnote}}
\renewcommand\footnoterule{\vspace*{1pt}%
\color{cream}\hrule width 3.5in height 0.4pt \color{black} \vspace*{5pt}} 
\setcounter{secnumdepth}{5}

\makeatletter 
\renewcommand\@biblabel[1]{#1}            
\renewcommand\@makefntext[1]%
{\noindent\makebox[0pt][r]{\@thefnmark\,}#1}
\makeatother 
\renewcommand{\figurename}{\small{Fig.}~}
\sectionfont{\sffamily\Large}
\subsectionfont{\normalsize}
\subsubsectionfont{\bf}
\setstretch{1.125} 
\setlength{\skip\footins}{0.8cm}
\setlength{\footnotesep}{0.25cm}
\setlength{\jot}{10pt}
\titlespacing*{\section}{0pt}{4pt}{4pt}
\titlespacing*{\subsection}{0pt}{15pt}{1pt}

\fancyfoot{}
\fancyfoot[LO,RE]{\vspace{-7.1pt}\includegraphics[height=9pt]{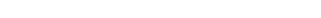}}
\fancyfoot[CO]{\vspace{-7.1pt}\hspace{13.2cm}\includegraphics{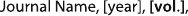}}
\fancyfoot[CE]{\vspace{-7.2pt}\hspace{-14.2cm}\includegraphics{head_foot/RF}}
\fancyfoot[RO]{\footnotesize{\sffamily{1--\pageref{LastPage} ~\textbar  \hspace{2pt}\thepage}}}
\fancyfoot[LE]{\footnotesize{\sffamily{\thepage~\textbar\hspace{3.45cm} 1--\pageref{LastPage}}}}
\fancyhead{}
\renewcommand{\headrulewidth}{0pt} 
\renewcommand{\footrulewidth}{0pt}
\setlength{\arrayrulewidth}{1pt}
\setlength{\columnsep}{6.5mm}
\setlength\bibsep{1pt}

\makeatletter 
\newlength{\figrulesep} 
\setlength{\figrulesep}{0.5\textfloatsep} 

\newcommand{\topfigrule}{\vspace*{-1pt}%
\noindent{\color{cream}\rule[-\figrulesep]{\columnwidth}{1.5pt}} }

\newcommand{\botfigrule}{\vspace*{-2pt}%
\noindent{\color{cream}\rule[\figrulesep]{\columnwidth}{1.5pt}} }

\newcommand{\dblfigrule}{\vspace*{-1pt}%
\noindent{\color{cream}\rule[-\figrulesep]{\textwidth}{1.5pt}} }

\makeatother

\twocolumn[
  \begin{@twocolumnfalse}
{\includegraphics[height=30pt]{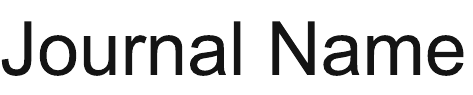}\hfill\raisebox{0pt}[0pt][0pt]{\includegraphics[height=55pt]{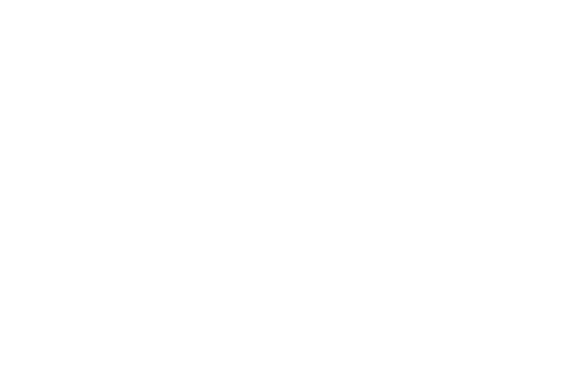}}\\[1ex]
\includegraphics[width=18.5cm]{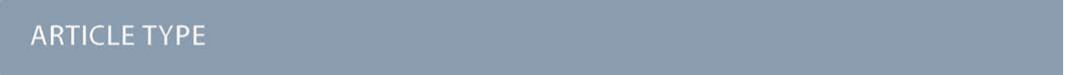}}\par
\vspace{1em}
\sffamily
\begin{tabular}{m{4.5cm} p{13.5cm} }

\includegraphics{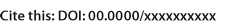} & \noindent\LARGE{\textbf{Long-time Brownian dynamics of hard and sticky colloids by geometric projection$^\dag$}} \\
 & \vspace{0.3cm} \\

& \noindent\large{
Rémi Martin,\textit{$^{a}$}
Sandrine Duluard,\textit{$^{a}$}
and Céline Merlet$^{\ast}$\textit{$^{a}$}
} \\

\includegraphics{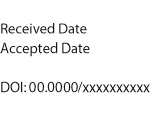} & \\

\end{tabular}

 \end{@twocolumnfalse} \vspace{0.6cm}

  ]

\renewcommand*\rmdefault{bch}\normalfont\upshape
\rmfamily
\section*{}
\vspace{-1cm}


\footnotetext{\textit{$^{a}$~Université de Toulouse, Toulouse INP, CNRS, CIRIMAT, Toulouse, France.}}
\footnotetext{$\ast$~E-mail: celine.merlet@utoulouse.fr}
\footnotetext{$\dag$~Electronic Supplementary Information (ESI) available:
methodological details, validation tests, performance benchmarks,
and additional details of the electrophoretic deposition simulations and analyses.}




\sffamily{\textbf{
Particle-resolved colloid simulations are often limited by the fastest microscopic time scale retained in the model, which can be many orders of magnitude shorter than the collective process of interest. This is especially restrictive for aggregating colloids, where explicit near-contact potentials force the time integration to resolve the rapid positional relaxation of bonded pairs within narrow attractive wells. Here we introduce an overdamped constraint-based Brownian dynamics method that replaces the time-resolved integration of stiff near-contact forces by a geometric projection step. After each free Brownian update, hard-sphere exclusion, sticky bonds, and hard-wall interactions are enforced by projecting the particle positions onto the constrained configuration set. The method preserves hard-sphere structure and diffusion, recovers the aggregation kinetics of sticky hard spheres, and extends the physical time reached at fixed computational cost by one to two orders of magnitude relative to the explicit-potential simulations tested here. This acceleration makes Brownian-dominated, low-Péclet electrophoretic deposition accessible to particle-scale simulations, revealing a pathway in which the suspension first densifies and only subsequently develops a bonded network. Implemented as a LAMMPS fix, the framework provides a modular route to long-time Brownian simulations of dense constrained colloidal systems.
}
}\\


\rmfamily 


For colloidal suspensions, Brownian dynamics (BD) provides a particle-scale description of how microscopic interactions drive collective phenomena such as aggregation, gelation, sedimentation and deposition. However, the accessible physical duration of these simulations is often set by the fastest time scale retained in the numerical model, leaving many slower processes of interest beyond the reach of a direct particle-resolved simulation. Brownian dynamics already removes one such time scale by eliminating inertial relaxation and retaining only the overdamped configurational dynamics\cite{ermak_brownian_1978}. Nevertheless, this does not eliminate all microscopic stiffness: explicit near-contact potentials can introduce short configurational relaxation times. A steep repulsive wall and a narrow attractive well differ in the shape of the interaction potential, but both generate rapid positional relaxation that must be resolved by the integrator. For collective observables on time scales much longer than near-contact relaxations, such as aggregation kinetics or structural correlations, the detailed relaxation pathway is often irrelevant: particles cannot overlap, and irreversibly adhesive contacts remain bonded. The corresponding limiting description is therefore naturally reformulated in terms of geometric constraints rather than time-resolved near-contact forces\cite{wertheim_exact_1963,baxter_percusyevick_1968}. For hard spheres, this idea underlies several potential-free Brownian schemes, from event-driven\cite{scala_event-driven_2007} and rejection-based\cite{cichocki_dynamic_1990} methods to overlap-correction\cite{heyes_brownian_1993,strating_brownian_1999} and reflected-transition-probability\cite{behringer_brownian_2012} approaches. At high density, however, their efficiency can be challenged by frequent encounters and many-body overlap events. This difficulty is further amplified by irreversible aggregation: sticky contacts persist over many Brownian steps and, in dense clusters, gels, or deposits, form strongly coupled constraint networks. Here we take a complementary route centered on a simple, well-defined geometric prescription designed to remain flexible and robust in dense, strongly coupled constraint networks. We formulate hard exclusion, sticky contacts, and wall interactions within a common constraint-projection framework for overdamped BD. The goal is to remove irrelevant fast modes associated with near-contact interactions while retaining longer-range smooth forces, Brownian diffusion, external driving, and the dense hard or sticky contact networks that control long-time aggregation and deposition pathways.
\medskip

Once stiff near-contact interactions are replaced by hard or sticky contact rules, their role in the dynamics shifts from providing forces to restricting the admissible domain of configuration space. We exploit this by enforcing these geometric constraints on the configuration obtained after each free BD step, borrowing from the computer-graphics framework of position-based dynamics\cite{muller_position_2007,macklin_xpbd_2016}, but integrated within an overdamped stochastic integrator. The instantaneous configuration of a system is described by the set of coordinates $\mathbf{X}$. Although the method can be applied to various types of degrees of freedom, such as rigid-body transformations, in this work we restrict ourselves to the Cartesian coordinates of the centers of $N$ spherical particles, $\mathbf{X}=\{\mathbf{r}_i\}_{i=1}^N$. 
The admissible configurations are expressed through a set of $N_\mathrm{cons}$ coupled constraint residuals, $\mathcal{C}=\{C_\alpha(\mathbf{X})\}_{\alpha=1}^{N_\mathrm{cons}}$, with $C_\alpha(\mathbf{X})=0$ for all satisfied constraints.
The interaction of a particle $i$ of radius $a_i$ with a lower hard wall is modeled by
\begin{equation}
    C_i^\mathrm{wall}(\mathbf{X}) = \max (0,\ z_i^0-z_i),
\end{equation}
where $z_i^0=z^\mathrm{wall}+a_i+h^0$ is the lowest $z$-coordinate allowed for the particle, with a minimum gap $h^0$ imposed on all the contacts.
We model the bonded interaction of particles $i$ and $j$ by
\begin{equation}
    C_{ij}^\mathrm{bond}(\mathbf{X})=d_{ij}^0-r_{ij},
\end{equation}
where $d_{ij}^0=a_i+a_j+h^0$, and $r_{ij}=\left\|\mathbf{r}_j-\mathbf{r}_i\right\|$.
Similarly, for non-bonded particles $i$ and $j$, we model their hard-sphere interaction by
\begin{equation}
    C_{ij}^\mathrm{HS}(\mathbf{X})=\max (0,\ d_{ij}^0-r_{ij}).
\end{equation}
With the discontinuous interactions encoded in $\mathcal{C}$, time integration alternates between a free BD update under smooth forces and a geometric projection.
The free update yields an unconstrained configuration $\mathbf{X}^{(0)}$, from which the projection solver constructs a sequence $\mathbf{X}^{(k)}$ until the maximum absolute residual falls below the tolerance $\epsilon$.
An iteration of the solver updates the configuration according to $\mathbf{X}^{(k+1)}=\mathbf{X}^{(k)}+\omega\,\Delta \mathbf{X}^{(k)}$, where \(\omega \leq 1\) is a relaxation factor and the unrelaxed correction is given by 
\begin{equation}
    \Delta \mathbf{X}^{(k)} = 
    - \sum_{\alpha}
    \frac{C_\alpha \left(\mathbf{X}^{(k)}\right)}{
    \nabla_\mathbf{X}C_\alpha\left(\mathbf{X}^{(k)}\right)^\mathrm{T}
    \cdot \mathsf{M}\cdot
    \nabla_\mathbf{X}C_\alpha\left(\mathbf{X}^{(k)}\right)
    }
    \mathsf{M}\cdot \nabla_{\mathbf{X}} C_\alpha\left(\mathbf{X}^{(k)}\right),
\end{equation}
where \(\mathsf{M}\) is the mobility matrix. 
In the present Cartesian representation, hydrodynamic interactions are neglected and \(\mathsf{M}\) is therefore a \(3N\times3N\) diagonal matrix, with each coordinate belonging to particle \(i\) assigned the Stokes mobility \(\mu_i=1/\gamma_i\).
All pair-constraint corrections are evaluated from $\mathbf{X}^{(k)}$, accumulated independently, and applied simultaneously.
This simultaneous update is naturally parallel, but can lead to persistent correction oscillations in dense bonded systems; under-relaxation with \(\omega<1\) was therefore used to improve convergence robustness.

\begin{figure}[b]
\centering
  \includegraphics[width=8.5cm]{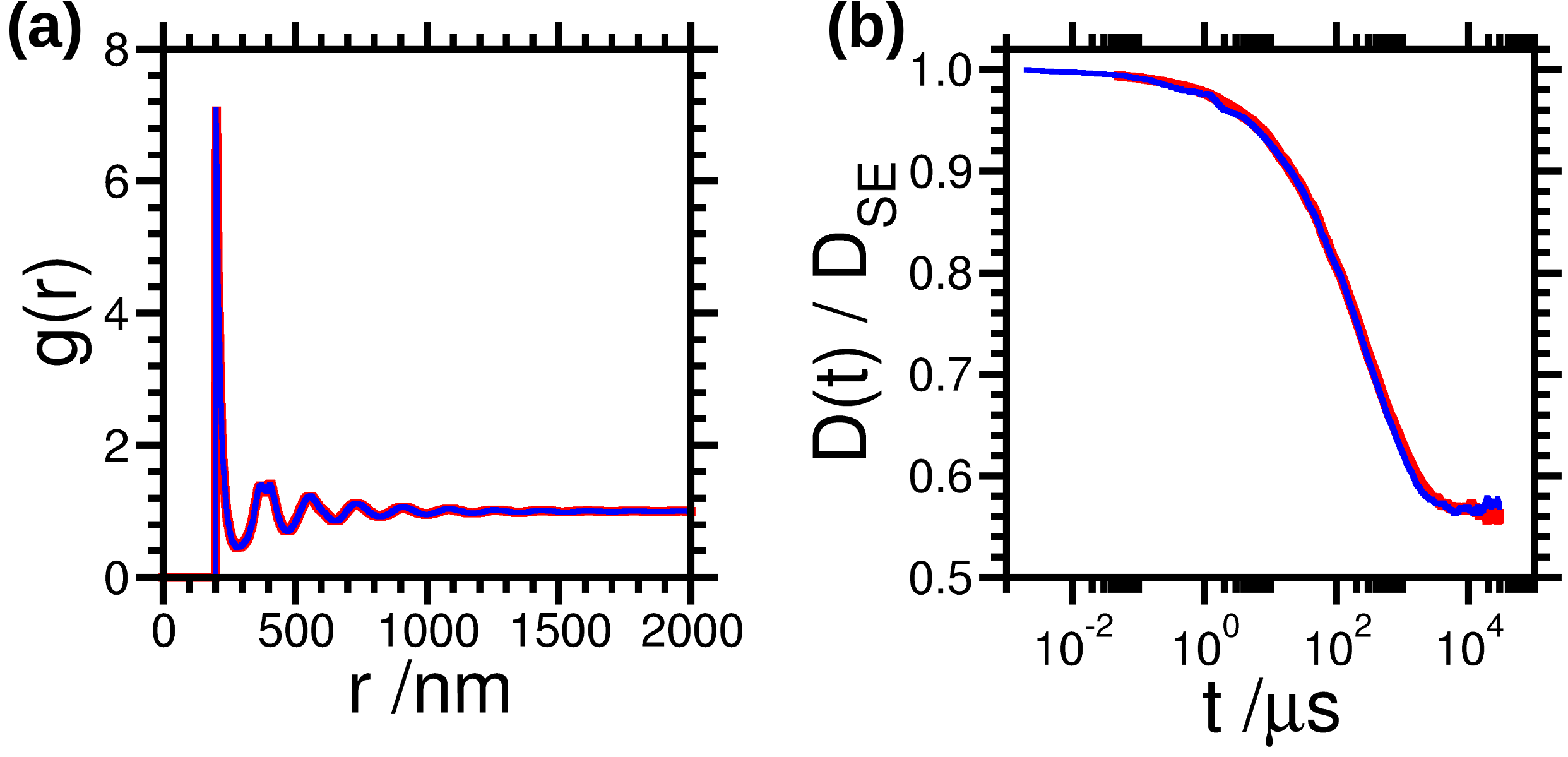}
  \caption{Validation of hard-sphere dynamics. (a) Radial distribution functions at $\phi=0.55$. (b) Time-dependent diffusion coefficient at $\phi=0.25$. Geometric hard-sphere projection (red) reproduces the reference steep-repulsion simulations (blue).}
  \label{fgr:msd_rdf}
\end{figure}
Adhesive aggregation is introduced as an update of the constraint set: a non-bonded pair is assigned an irreversible bond constraint when $r_{ij}<d_{ij}^0+\epsilon$.
The same $\epsilon$ is used as the projection tolerance, ensuring consistency between bond creation and bonded-contact projection.
Together, the contact distance $h^0$ and the tolerance $\epsilon$ set the geometric resolution of near-contact dynamics and are chosen to balance stable convergence with physically consistent long-time behavior.
The resulting scheme explicitly separates smooth interactions, which generate the free Brownian step, from singular hard and sticky interactions, which define the admissible geometry onto which the configuration is projected.
The method is implemented as a LAMMPS fix\cite{LAMMPS-basic,thompson_lammps_2022} and is available, with input examples, in a zenodo archive (DOI: 10.5281/zenodo.23802331). Further details of the Brownian update, constraint projection, and bond handling are provided in the ESI$^\dag$.
\medskip

\begin{figure}[h]
\centering
  \includegraphics[width=8.5cm]{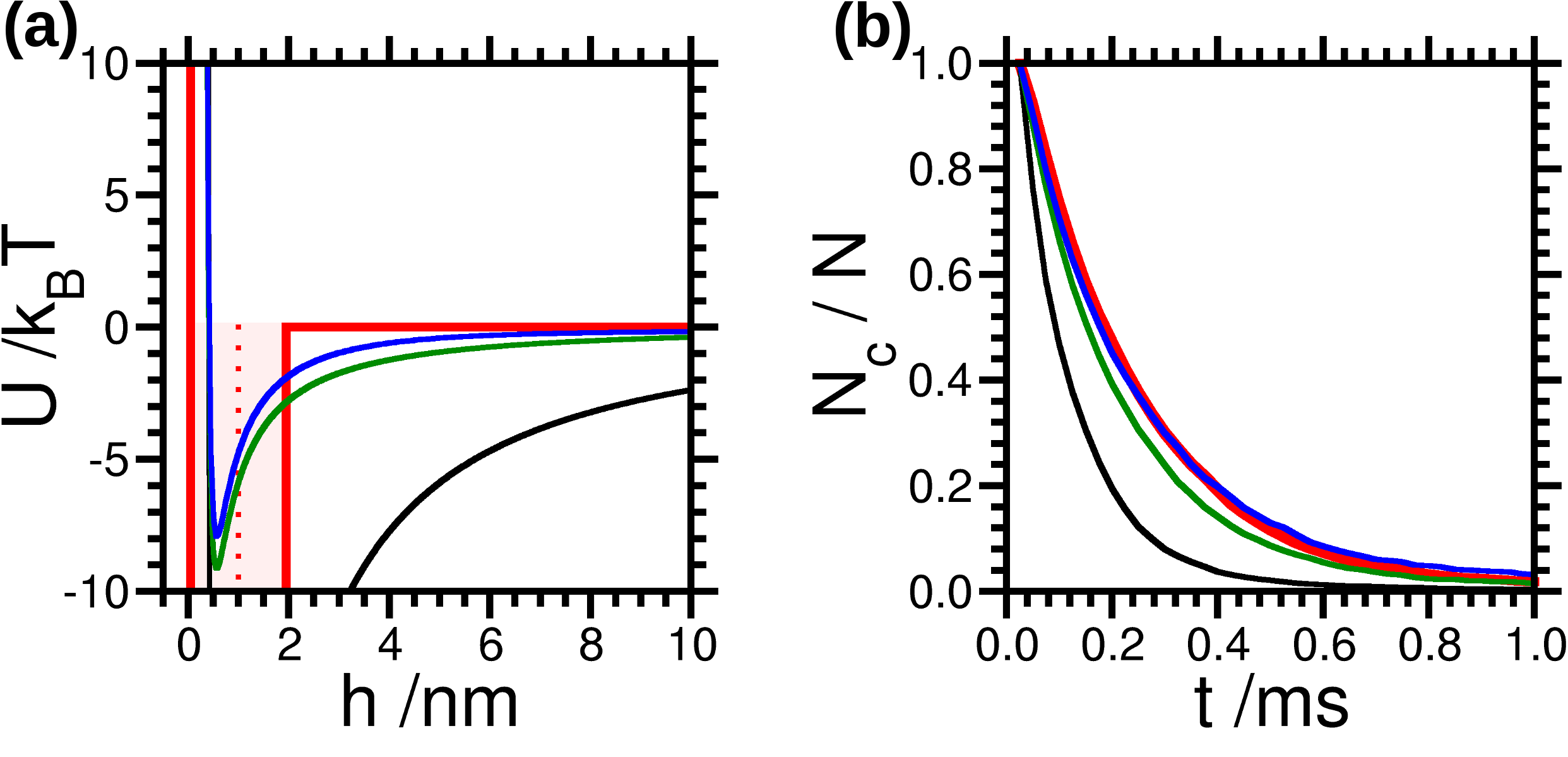}
  \caption{Validation of aggregation kinetics.
  (a) Aggregating pair potentials with decreasing attractive range: potential 1 (black), potential 2 (green), and potential 3 (blue), compared with the idealized infinite square-well representation of the projected sticky-contact model (red). 
  (b) Time evolution of the normalized cluster number, $N_c/N$, for the same models at \(\phi=0.25\).
  The projected sticky hard-sphere model reproduces the aggregation kinetics of the shortest-range reference potential.}
  \label{fgr:aggreg_kinetics}
\end{figure}
We validate the constraint formulation in the two elementary limits on which the method relies: non-aggregating hard-sphere Brownian motion and irreversible sticky aggregation.
In the non-aggregating case, the projection should enforce excluded volume without altering either the equilibrium structure or the long-time diffusive dynamics.
We therefore compare hard-sphere constrained simulations with explicit-potential BD that uses the repulsive branch of the integrated Lennard-Jones potential\cite{everaers_interaction_2003}, truncated at its minimum.
The radial distribution function is measured at $\phi=0.55$, where steric structure is strongly constrained, while the time-dependent diffusion coefficient $D(t)$ is measured at $\phi=0.25$, where the long-time diffusive regime remains accessible to explicit-potential BD.
As shown in Fig.~\ref{fgr:msd_rdf}, the projected dynamics preserves the tested equilibrium structure and diffusive dynamics within the tested time-step range.
We then test the adhesive contact rule at $\phi=0.25$ by comparing the normalized cluster number $N_c(t)/N$ to explicit BD simulations with attractive integrated Lennard-Jones potentials of different ranges and well shapes, potential 1, 2 and 3, shown in Fig.~\ref{fgr:aggreg_kinetics}.
These explicit potentials retain the near-contact attraction but still require the small time step associated with fast bonded-pair relaxation.
As the attraction becomes short-ranged but still effectively irreversible, the explicit-potential kinetics approaches the sticky-contact result, showing that the constraint rule captures the contact-limited aggregation dynamics targeted by the coarse-grained model.
These comparisons were also used to select the largest time step for which the relevant observables remained unchanged, thereby defining the validated time steps used in the performance measurements below.
Time-step sensitivity tests and complete simulation parameters are reported in the ESI$^\dag$.

\begin{figure}[h]
\centering
  \includegraphics[width=8.5cm]{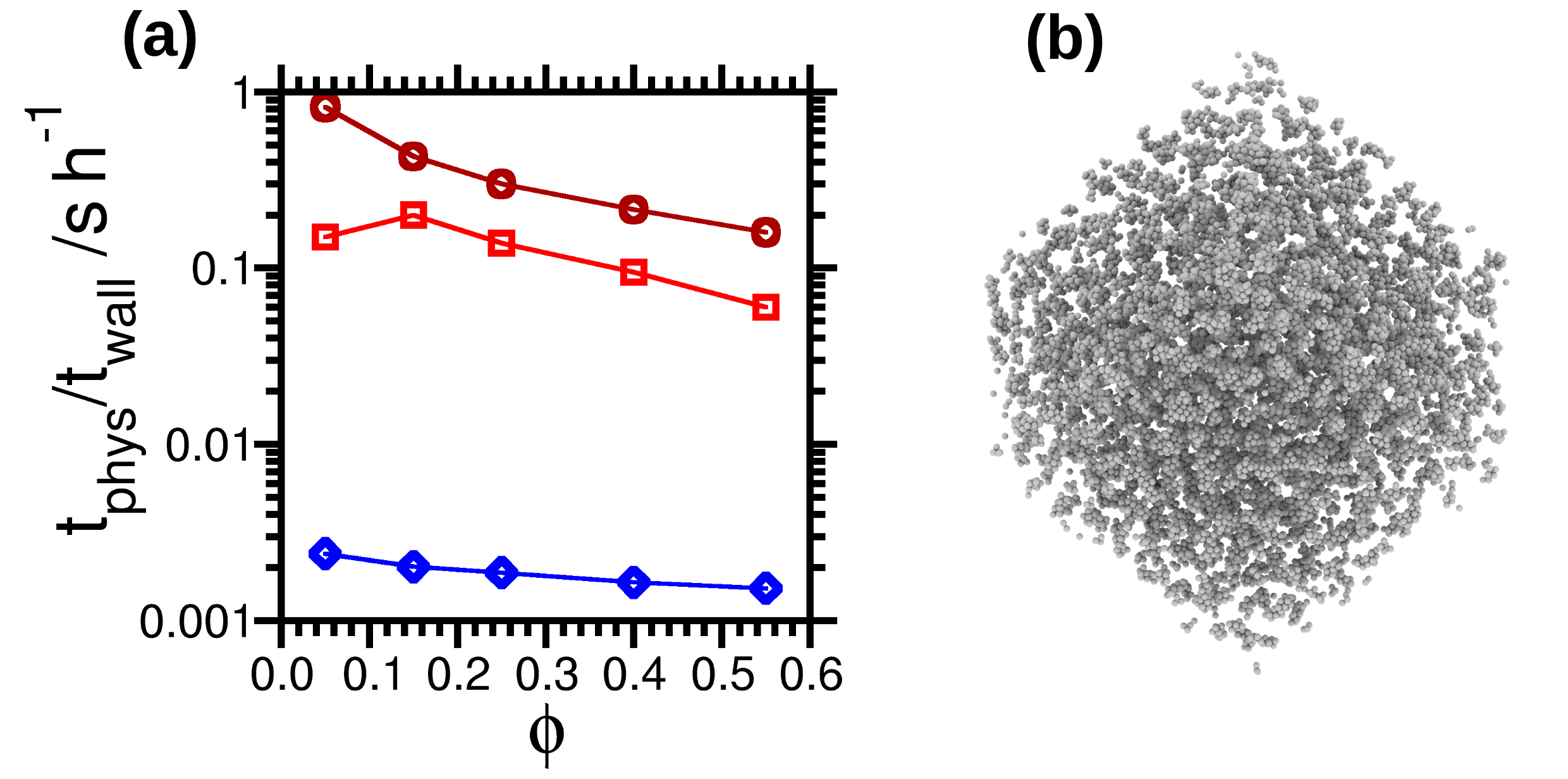}
  \caption{
  (a) Physical-time throughput, $t_\mathrm{phys}/t_\mathrm{wall}$, as a function of volume fraction $\phi$. 
  Explicit-potential BD with the repulsive branch of the integrated Lennard-Jones potential truncated at its minimum is shown in blue diamonds; projected hard-sphere BD in red circles; and projected sticky aggregates, initialized with an average cluster size of 10 particles, in red squares. 
  (b) Snapshot of an initial aggregate configuration generated at $\phi=0.05$ for the sticky-aggregate benchmarks.
  }
  \label{fgr:performance}
\end{figure}

The validated time steps already reveal the origin of the computational gain: explicit-potential BD required $\Delta t=50\,\mathrm{ps}$ for non-aggregating suspensions and $\Delta t=25\,\mathrm{ps}$ for aggregating ones, whereas the constraint-based simulations reached $\Delta t=50\,\mathrm{ns}$ for both non-aggregating hard spheres and sticky hard spheres.
Because this three-order-of-magnitude increase in physical time per step comes with the additional cost of completing the projection, we assess the net speed increase through physical time throughput $t_\mathrm{phys}/t_\mathrm{wall}$ where $t_\mathrm{phys}$ is the simulated physical time and $t_\mathrm{wall}$ is the corresponding wall-clock time.
Figure~\ref{fgr:performance}a shows a throughput one to two orders of magnitude larger than explicit-potential BD over the tested volume fractions, despite the increasing projection cost at high density and in bonded systems.
The solver was deliberately chosen for its simplicity and parallel implementation: constraint corrections are evaluated independently before being applied simultaneously.
This design is naturally parallel but can require many iterations to converge, especially in dense bonded networks.
Thus, even with this simple projection scheme, the gain obtained by removing the stiff near-contact time scale is not lost to the solver.
This computational margin is what makes it possible to address low-field deposition regimes that remain impractical with explicit potentials.
Benchmark definitions, simulation parameters, and parallel-scaling results are provided in the ESI$^\dag$.
\medskip

\begin{figure*}
 \centering
 \includegraphics[width=18.5cm]{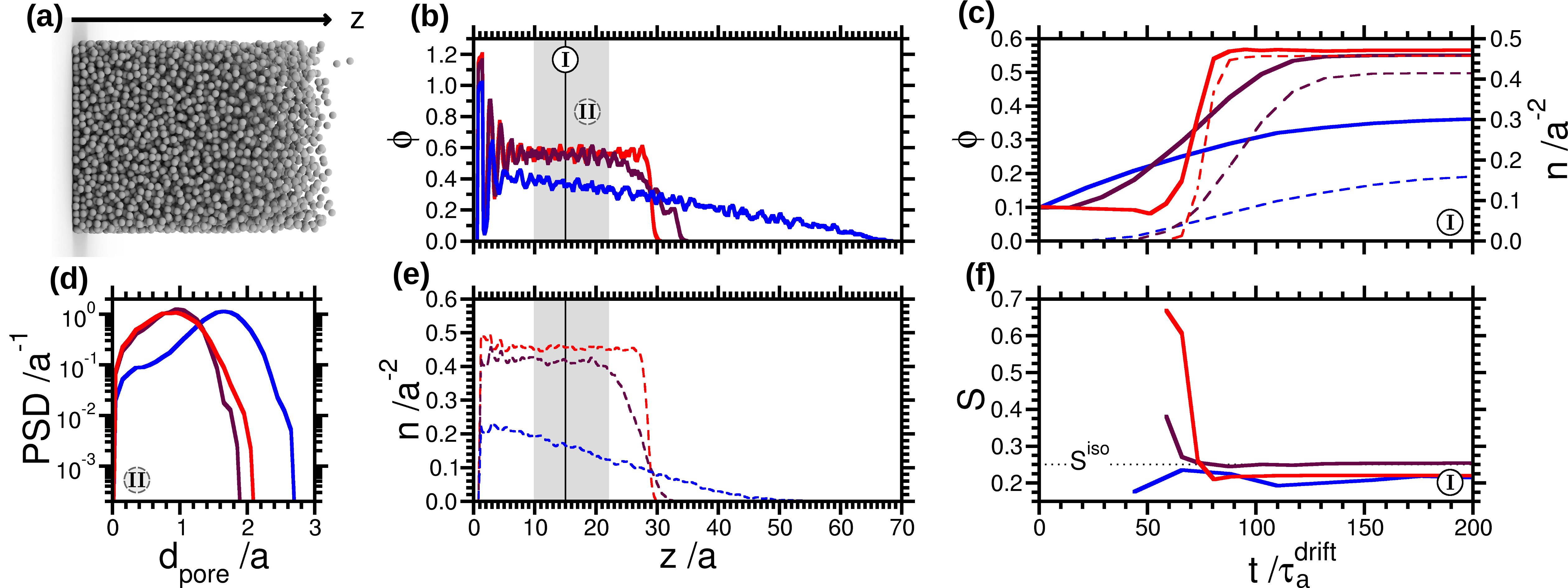}
 \caption{Low-Péclet EPD displays a deposition pathway different from high-Péclet EPD, where densification occurs prior to the formation of a bonded network. 
 (a) Snapshot of the final deposit obtained at $\mathrm{Pe}=1$.
 The remaining panels show properties resulting from one EPD simulation at each Péclet number: $\mathrm{Pe}=1$ (blue), $\mathrm{Pe}=10$ (purple), and $\mathrm{Pe}=100$ (red).
 (b) Final particle density profiles along the deposition direction, expressed as volume fractions.
 (e) The corresponding final profiles of the surface density of bonds crossing planes normal to $z$.
 (d) Final pore-diameter distributions in the deposit core, averaged over the grey region indicated in (b) and (e).
 The deposition dynamics are shown in (c) and (f).
 (c) Time evolution at $z=15a$ of the local volume fraction $\phi$ (solid lines) and plane-crossing bond surface density $n$ (dashed lines).
 (f) Time evolution at $z=15a$ of the bond-orientation order parameter $S$, computed from bonds crossing the plane normal to $z$.
 In (c) and (f), time is reported in units of the particle-scale drift time $\tau^\mathrm{drift}_a=a/|v^\mathrm{drift}|$.}
 \label{fgr:EPD}
\end{figure*}

To demonstrate the new dynamical regimes made accessible by the method, we apply it to electrophoretic deposition (EPD), a coating technique in which charged particles are driven by an electric field toward an electrode, where densification and irreversible coagulation jointly lead to deposit formation\cite{sarkar_electrophoretic_1996,besra_review_2007}.
Despite extensive study, direct experimental access to particle-scale dynamics during EPD remains limited, particularly in concentrated suspensions and growing deposits, which complicates mechanistic interpretation and makes particle-resolved simulations valuable for linking prescribed effective interactions to deposition pathways\cite{corni_electrophoretic_2008}.
Lowering the electric field directly lowers the Péclet number,
\begin{equation}
    \mathrm{Pe} = \frac{|v^{\mathrm{drift}}|a}{D}
    = \frac{|q_\mathrm{eff}E|a}{k_\mathrm{B}T},
\end{equation}
and increases the drift time over a distance $L$,
\begin{equation}
    \tau^\mathrm{drift}_L
    = \frac{L}{|v^\mathrm{drift}|}
    = \frac{1}{\mathrm{Pe}}\frac{aL}{D}.
\end{equation}
Therefore, at fixed drift-scaled deposition time, lowering $\mathrm{Pe}$ increases the physical duration to be simulated, while explicit DLVO Brownian dynamics remains constrained by the short time step imposed by the stiff near-contact potential.
A finite wall-clock budget therefore sets a lower bound on the Péclet numbers that can be reached. 
In a favorable setup, with $N=10,000$ particles, initial volume fraction $\phi=0.1$ and box length $L_z=167a$ along the field direction, explicit DLVO simulations would already require several weeks to reach $\mathrm{Pe}=10$ and would remain impractical at $\mathrm{Pe}=1$. 
In contrast, the present constraint-based method reaches $\mathrm{Pe}=10$ in about 1.5 days and $\mathrm{Pe}=1$ in about 6 days on the same computational setup, enabling the comparison of the deposition pathways shown in Fig.~\ref{fgr:EPD}.
The effective interaction, simulation setup, and time-step validation for the EPD calculations are detailed in the ESI$^\dag$.

To characterize the local formation of the bonded network, we measure the surface density $n(z)$ of bonds crossing a plane at height $z$, and their orientational order parameter $S(z)$ along the deposition direction,
\begin{equation}
    S(z)
    =
    \frac{
    \sum_{\langle i,j\rangle_z}
    \left(\frac{3}{2}u_{ij,z}^2-\frac{1}{2}\right)
    }{
    \sum_{\langle i,j\rangle_z}1
    },
    \label{eq:bond_orientation_order}
\end{equation}
where $u_{ij,z}$ is the $z$-component of the bond unit vector, and $\langle i,j\rangle_z$ denotes bonds intersecting the plane at height $z$.
Because this plane-crossing selection weights bonds by their probability of intersecting the plane, an isotropic bond network gives $S^\mathrm{iso}=1/4$.
Details of the density, bond-orientation, and pore-size analyses are provided in the ESI$^\dag$.
The accessible low-$\mathrm{Pe}$ regime already displays signatures that are not captured by a simple extrapolation from high-field deposition.
At high $\mathrm{Pe}$, particles are rapidly driven into the growing front: at a given height, the local volume fraction and the surface density of crossing bonds rise sharply, and the first bonds are preferentially oriented along the field direction.
Accordingly, $S(t)$, measured at $z=15a$, starts above the isotropic reference before relaxing toward it as the deposit densifies and rearranges.
At $\mathrm{Pe}=1$, by contrast, the local density builds up more progressively and bond formation starts close to the isotropic reference, consistent with barrier-limited coagulation occurring upon densification rather than immediate front-induced bonding.
This change in deposition pathway also leaves a structural signature in the final deposit: the core pore-size distribution shifts from small interstitial voids at high $\mathrm{Pe}$ toward a larger population of particle-scale pores at $\mathrm{Pe}=1$.
These dynamical and structural signatures show that the method opens the way to systematic particle-resolved studies of low-field EPD regimes that were previously impractical with explicit potentials.
\medskip
       
Overall, these results show that stiff near-contact interactions can be removed from time integration when their relevant effect can be represented by geometric constraints and contact-capture rules.
The constraint-based Brownian dynamics scheme preserves the targeted hard-sphere and contact-limited aggregation observables while extending the accessible physical time, as illustrated here by low-$\mathrm{Pe}$ EPD.
Beyond the spherical hard and sticky contacts considered in this work, the formulation is modular: additional degrees of freedom and constraint functions, for example for non-spherical particles, can in principle be introduced within the same projection framework.
Extensions to reversible contacts or configuration-dependent hydrodynamic mobilities would, however, require additional ingredients beyond the present formulation.
More broadly, the approach provides a practical way to align the numerical time scales of Brownian dynamics simulations with the collective time scales of interest.

\newpage
\section*{Author Contributions}

Rémi Martin: Methodology, Software, Validation, Investigation, Formal analysis, Writing – Original Draft, Writing – Review \& Editing. Sandrine Duluard: Methodology, Writing – Review \& Editing, Supervision, Resources, Funding acquisition. Céline Merlet: Conceptualization, Formal analysis, Methodology, Validation, Writing – Review \& Editing, Supervision, Resources, Funding acquisition.

\section*{Conflicts of interest}

There are no conflicts to declare.

\section*{Data availability}

The Zenodo archive (\mbox{\href{https://doi.org/10.5281/zenodo.23802331}{DOI:~10.5281/zenodo.23802331}}) provides the LAMMPS implementation of the geometric projection method together with example input files for the simulations.

\section*{Acknowledgment}
The authors thank the French «Ministère de l’Enseignement Supérieur et de la Recherche et de l'Espace (MESRE)» for the PhD grant of Rémi Martin. This work was granted access to the HPC resources of the CALMIP supercomputing centre under the allocation P21014. The authors acknowledge Florence Ansart, Pierre-Louis Taberna, Vincent Dahirel and Marie Jardat for useful discussions.







\bibliography{rsc} 
\bibliographystyle{rsc} 

\end{document}